\documentclass[prl,twocolumn,superscriptaddress,showpacs]{revtex4}
\def\beq{\begin{equation}}
\def\eeq{\end{equation}}
\def\beqn{\begin{eqnarray}}
\def\l{\left}
\def\rr{\right}
\def\be{\begin{equation}}
\def\ee{\end{equation}}
\def\H{\mathcal{H}}
\def\summ{\sum\limits}
\def\prodd{\prod\limits}
\def\bra#1{\l\langle #1 \rr|}
\def\ket#1{\l| #1 \rr\rangle}
\def\eeqn{\end{eqnarray}}
\usepackage{amsmath}
\usepackage{graphicx}
\usepackage{epsfig}
\usepackage[usenames]{color}
\usepackage{subfigure} 
\begin{document}

\title{Majorana fermion chain at the Quantum Spin Hall edge}

\author{V. Shivamoggi}
\affiliation{Department of Physics, University of California, Berkeley}

\author{G. Refael}
\affiliation{Department of Physics, California Institute of Technology}

\author{J. E. Moore}
\affiliation{Department of Physics, University of California, Berkeley}
\affiliation{Materials Science Division, Lawrence Berkeley National Laboratory}

\begin{abstract}
We study a realization of a 1d chain of Majorana bound states at the
interfaces between alternating ferromagnetic and superconducting
regions at a quantum spin Hall insulator edge.  In the limit of
well separated Majoranas, the system can be mapped to the
transverse field Ising model.  The disordered critical point can be
reached by tuning the relative magnitude or phases of the ferromagnetic
and superconducting order parameters.  We compute the voltage
dependence of the tunneling current from a metallic tip into the
Majorana chain as a direct probe of the random critical state.
\end{abstract}

\maketitle


Some electronic phases support emergent quasiparticle excitations that
are in a sense fractions of the original electrons, as in the
fractional quantum Hall effect.  One of the most basic fractional
excitations is the Majorana fermion familiar from particle
physics: the Majorana fermion is its own antiparticle and represents
``half'' of an ordinary (Dirac) fermion.
Several condensed matter systems are 
theoretically believed to realize Majorana fermions in their excitation
spectrum, and their direct observation is a major goal of
current research. In this work we compute the tunneling conductance and other
experimental signatures of one of the simplest systems of many
Majorana fermions.


Proposals to observe Majorana
fermions~\cite{fukaneprox2d,fukaneprox3d,nilsson}  have been 
based on the proximity effect between an ordinary $s$-wave
superconductor and the recently discovered ``topological insulator''
phases in two-dimensional~\cite{km2,zhangscience1,molenkampscience}
and
three-dimensional~\cite{moore&balents-2006,rroy3D,fu&kane&mele-2007,hsieh}
materials.  These 
phases generally support
gapless edge or surface states.  In the two-dimensional (2d) case, the
gapless edge can be viewed as counterpropagating up-spin and down-spin
electrons, hence the terminology ``quantum spin Hall'' (QSH) for this
phase.  In the three-dimensional ``strong topological
insulator''~\cite{fu&kane&mele-2007}, the surface state is a
two-dimensional metal with a Fermi surface that encloses an odd number
of Dirac points.  The proximity effect between an ordinary
superconductor and this surface state leads to a superconducting
state~\cite{fukaneprox3d} that is time-reversal symmetric but
topologically similar to the $p+ip$ superconductor~\cite{readgreen},
with Majorana fermions trapped in vortices.


\begin{figure}
\subfigure[]
{
\includegraphics[width=0.4\textwidth]{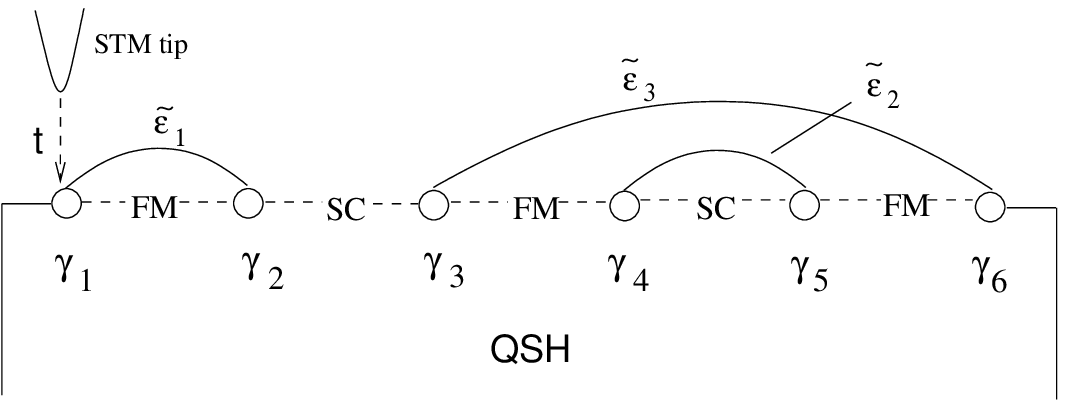}
\label{chaindec}
}

\subfigure[]
{
\includegraphics[width=0.3\textwidth]{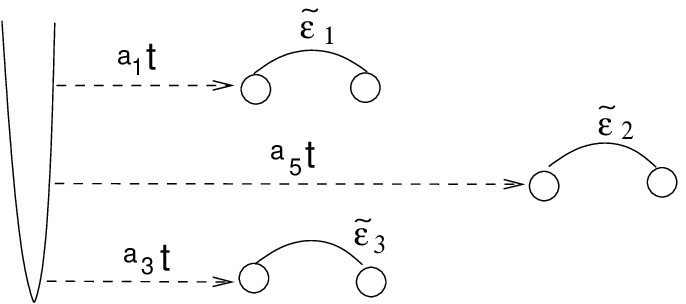}
\label{multpairs}
}
\caption{(a) A Majorana chain realization: An array of alternating FM and SC regions at the QSH edge.  Following a real-space RG
  analysis, the chain decouples into pairs, shown by the upper lines.
  $\tilde{\epsilon_i}$ is the energy of the pair decimated with the
  i-th highest energy.  (b) The RG effectively couples the STM tip
  to every Majorana pair, with tunneling suppressed by the
  coupling coefficient $a$.  Majorana pairs decimated at the edge are
  coupled more strongly to the lead than pairs decimated in the bulk
  of the chain. }
\end{figure}

Perhaps the most appealing from a technical perspective are Majorana
realizations in 1d systems. In particular, a QSH edge supports a
localized Majorana fermion excitation~\cite{kitaev2000,fukaneprox2d}.
When the edge state is gapped by coupling to a superconductor (SC) in
one region and a ferromagnet (FM) in another, a single zero-energy
Majorana fermion appears at the SC/FM boundary if it is sufficiently
narrow (Fig.~\ref{chaindec}).   In this paper we study how unique
features of a chain of Majorana fermions, created by alternately FM
and SC regions along a QSH edge, can be observed using a metallic
tunneling tip.

An ideal, uniform, system realizes the interesting free-Majorana
fermion quantum critical point. Realistically, however, FM and SC
domains will inevitably vary randomly in size and proximity amplitude,
and therefore the proposed system would realize an even more
interesting model: the random Majorana-fermion chain in its
random-singlet phase \cite{bonesteelyang}.  Previously, likely experimental
Majorana signatures were discussed for one or two Majorana pairs \cite{nilsson,bolechdemler, tanakaprl, asanotanaka}, and the phase diagram of a 2d Majorana fermion model without disorder was obtained~\cite{xufu}.  Experimentally, it
may be even simpler to pattern an irregular phase-separated mixture of
SC and FM materials to contact the QSH edge, rather than a precise
controlled-lithography FM-SC-FM configuration.

Furthermore, the Majorana fermion chain may give a first convincing
measurement of the random-singlet phase and its associated Griffiths
scaling. Contrary to spin chains where neutron scattering or other
spin-sensitive probes can only give limited information, the Majorana
chain setup allows a direct electronic tunneling measurement that
reveals the nature of the random singlet phase.  In what follows we
will first derive the Hamiltonian of the random Majorana chain
contacted by a tunneling tip, and then calculate its main tunneling
I-V features.


The Majorana chain Hamiltonian arises as the low energy (in-gap)
sector of the QSH edge Hamiltonian in proximity to SC and FM domains
alternately, with interfaces arising at $x_1=0$, and $x_i>0$
thereafter. The QSH edge Hamiltonian has the form~\cite{fukaneprox3d}
$H=\frac{1}{2}\Psi ^{\dag}\mathcal{H} \Psi$, where $\Psi=\left (  \psi
_{\uparrow}, \psi _{\downarrow}, \psi _{\downarrow} ^{\dag}, -\psi
_{\uparrow} ^{\dag}   \right ) ^T$ and
\beqn
\mathcal{H} = &-iv\partial _x \sigma ^z \tau ^z + \vec{\Delta} (x) \cdot  \vec{\tau} + \vec{M}(x) \cdot \vec{\sigma}  
\label{hamiltonian}
\eeqn
The first term describes the counterpropagating spins of the QSH edge, while the rest describes the proximity to FM regions ($i=2j$) and SC regions ($i=2j+1$) with $\vec{M}(x) = (M_0 \cos \phi _{mi}, M_0 \sin \phi _{mi}, 0)$ and
$\vec{\Delta}(x) = (\Delta _0 \cos \phi_{si}, \Delta _0 \sin \phi_{si}, 0)$. $\sigma^{x,y,z}$ are Pauli matrices that act in the space of right and left movers $\psi _{\uparrow}$ and $\psi _{\downarrow}$, while $\tau ^{x,y,z}$ are Pauli matrices that act on the $\psi$ and $\psi ^{\dag}$ blocks. For simplicity, we set all magnetic moments $M_0$ and
pairing strengths $\Delta_0$ uniform, but let the $x$-$y$ moment
directions $\phi_{M2j}$ and SC order parameter phase $\phi_{S2j+1}$
vary.  Magnetic moment along $z$ does not open a gap and is irrelevant
for what follows. Each interface at $x_i$ will host a Majorana state, whose creation
operator we will denote $\gamma_i$. Unlike standard fermion creation
operators, a Majorana fermion obeys $\gamma^{\dagger}=\gamma$ and
$\gamma^2=1$. Two Majorana fermions can combine to give a Dirac
fermionic state via the transformation
$c=\frac{1}{2}(\gamma_1+i\gamma_2),\,c^{\dagger}=\frac{1}{2}(\gamma_1-i\gamma_2)$;
  a pair-wise Majorana interaction is simply written as:

\be
\H_{12}=-it_{12}\gamma_1\gamma_2=t_{12}(2 c^{\dagger}c-1).
\ee
In a Majorana chain such an interaction weakly couples nearest neighbor states. 

It is instructive to derive this interaction for just two Majoranas located at $x_1=0$ and $x_2=L$ due to a FM-SC-FM sequence, described by $\phi_{M0},\,\phi_{S1},\,\phi_{M2}$ respectively. To obtain the Majorana state at $x_1=0$ we ignore the $x_2$ interface and find a zero energy Majorana solution of the Hamiltonian (\ref{hamiltonian}) localized about $x=0$:
\beqn
\xi_1(x)=e^{\frac{i}{2}\sigma_z\phi_{M0}+\frac{i}{2}\tau_z\phi_{S1}}\sqrt{\frac{2v
    M_0\Delta_0}{M_0+\Delta_0}} e^{-|x|/\ell_{\pm}}\l(1,-i,i,-1\rr)^T
\eeqn
with the decay lengths for $x>0$ and $x<0$ being $\ell_+=v/\Delta_0$ and $\ell_-=v/M_0$ respectively. The Majorana
creation operator is thus $\hat{\gamma_1}=\int dx \xi_{1}(x)\Psi(x)$. Duality yields the second Majorana state at $x_2=L$:
an exchange $\Delta$ and $M$ in Eq. (\ref{hamiltonian}) switches the second and third components $\xi(2)\leftrightarrow\xi(3)$ in the resulting wave function vector. Thus the $x_2=L$ Majorana is: 
\beqn
\xi_2(x)=e^{\frac{i}{2}\sigma_z\phi_{M2}+\frac{i}{2}\tau_z\phi_{S1}}\sqrt{\frac{2v
    M_0\Delta_0}{M_0+\Delta_0}} e^{-x/\ell_{\pm}}\l(1,i,-i,-1\rr)^T
\eeqn
with $\ell_+=v/M_0$ for $x>L$, and $\ell_-=v/\Delta_0$ for $x<L$. Once again we define $\hat{\gamma_2}=\int dx \xi_{2}(x)\Psi(x)$. 

The Majorana wave function $\xi_1$ is a zero-energy eigenstate of (\ref{hamiltonian}) ignoring all interfaces
other than $x_1$. The interface at $x_2$ clearly gives rise to hybridization of $\gamma_1$ with $\gamma_2$. This we obtain by
writing  $\H=\H_0+V$ with $\H_0$ describing just the FM/SC interface at $x_1$ and:
\beqn
V=\Theta(x-L)(\vec{\Delta}_1 \cdot \vec{\tau}_{xy}-\vec{M}_2 \cdot \vec{\tau}_{xy}).
\eeqn
$\Theta(x)$ is the Heaviside function. The perturbation piece $V$ produces the hybridization
\beqn
-it_{12}=\bra{\xi_1}V\ket{\xi_2}=\int dx \xi_1^{\dagger}
V\xi_2(x).
\eeqn
Writing this hybridization so that it captures SC/FM/SC as well as FM/SC/FM Majorana pairs, we obtain
\be
t_{i,i+1}=2\cos\l(\frac{\Delta\phi_i}{2}\rr)\frac{M_0\Delta_0}{\Delta_0+M_0}e^{-L/\ell_i}.
\label{teq}
\ee
with $\Delta\phi_{2j+1}=\phi_{M2j+2}-\phi_{M2j}$ and $\ell=v/\Delta_0$ describing a FM/SC/FM sequence, and
$\Delta\phi_{2j}=\phi_{S2j-1}-\phi_{S2j+1}$ and $\ell=v/M_0$ for a
SC/FM/SC sequence. Since the interaction between Majorana fermions
decays exponentially with distance, we keep only the nearest neighbor
interaction, and the Hamiltonian of the chain becomes $\sum_i t_{i,
  i+1} \gamma _i \gamma _{i+1}$.  The strong-disorder random-singlet fixed point is
attained when
$\overline{\ln t_{2j, 2j+1}} = \overline{\ln t_{2j+1, 2j+2}}$~\cite{DSF95}.

Mapping this Hamiltonian to the quantum Ising spin chain~\cite{bonesteelyang} immediately yields critical thermodynamic~\cite{DSF95} and entanglement~\cite{refaelmoore} properties of the Majorana chain, which seem difficult to measure in practice.  Instead we focus on the local density of states, which is probed directly using a single STM lead biased at voltage $V$ relative to the chain. The combined Hamiltonian for the Majorana chain and a probing tip coupled to one end is:
\beq
H = -i\summ_{i=1}^{\infty}t_{i,i+1}\gamma_i\gamma_{i+1} + it_0\gamma_1
\left( \psi ^{\dag}e^{-ieVt/\hbar} + \psi e^{ieVt/\hbar}\right)
\label{leadham}
\eeq
where $t_{i,i+1}$ is given in Eq.~\ref{teq}, and $\psi$ is the
annihilation operator of tip electrons ~\cite{bolechdemler}.  We will
use real-space RG to reduce the semi-infinite chain and STM to
aproblem of a tip interacting with independent Majorana pairs:
\be
\H_{eff}\approx -i\summ_{n=1}^{\infty} (a_n\gamma_{n,a} \left( \psi ^{\dag} e^{-ieVt/\hbar} + \psi e^{ieVt/\hbar}\right)+\epsilon_n\gamma _{n, a}\gamma_{n,b}).
\ee
The decoupled Majorana pairs correspond to the so-called random-singlets formation that was investigated in random Heisenberg chains \cite{DSF94} and in random hopping fermions \cite{GirvinYang}. The I-V characteristics of this Hamiltonian are clear. Whenever the voltage of the tip is at resonance with $\pm\epsilon_n$ a step appears in the $I-V$ curve, with an amplitude proportional to the tunneling rate, $a_n^2$:
\be
\frac{dI}{dV}\sim
\summ_n a_n^2(\delta(eV+\epsilon_n)-\delta(eV-\epsilon_n)).
\label{LDOS}
\ee
The two opposite sign delta-functions are due to the special coupling term $\propto\psi+\psi^{\dagger}$; if at a positive resonance, $V=|\epsilon_n|$, $\psi$ is responsible for a current flowing {\it into} the chain, and when
$V=-|\epsilon_n|$, $\psi^{\dagger}$ will produce the current, which flows {\it out of} the chain to the tip. 

The measured LDOS, Eq. (\ref{LDOS}), will provide not only the chain's
density of states, but also their overlaps $a_n$ with the edge
Majorana. Therefore it reveals both the energy and
spatial structure of the random Majorana chain. 
We can find both the energy levels $\epsilon_n$ and the edge overlap,
$a_n$ using the real space RG method applied as a series of unitary
transformations on Eq.~\ref{leadham} ~\cite{Refael2003}. Its advantage
is that it does not eliminate pieces of the Hilbert space, but rather
decomposes the Hamiltonian Eq. (\ref{leadham}) into decoupled Majorana
pairs.


The unitary real-space RG procedure finds a sequence of
unitary transformations which isolate the strongest bonds, at each
stage, in Eq. (\ref{leadham}) from the rest of the chain. For
instance, if bond $n$ is the strongest, we find that using
$S^{(n)}=\frac{t_{n-1}}{2t_n}i
\gamma_{n-1}\gamma_{n+1}-\frac{t_{n+1}}{2t_n}i \gamma_{n}\gamma_{n+2}$
yields:
\be
e^{iS^{(n)}}\summ_j \H_j e^{-iS^{(n)}}=\summ_{j\neq{n\pm 1}}
  \H_j -i\frac{t_{n-1}t_{n+1}}{2t_n}\gamma_{n-1}\gamma_{n+2}
\ee
to order $O\l(\frac{1}{t_n^2}\rr)$. Thus, bond $n$ is decoupled, and
bridging over it, a
new coupling between sites $n-1$ and $n+2$ appears. Iteratively applying the transformations $S_i$, which decouple the sites $n_{ia}$ and $n_{ib}$,
also yields a wave function for the interacting chain, which is just a
transformed product wave function of the decoupled pairs:
\be
\ket{\Psi}=\prodd_j e^{-iS_j}\prodd_i\ket{\Psi_{n_{ia}n_{ib}}}.
\ee

To find the LDOS, we need to compute ground state matrix elements of
the contact term in Eq. (\ref{leadham}). Our method allows us to write
such matrix elements in terms of the decoupled Majorana-pairs wave
functions, $\ket{\Psi_{n_{ia}n_{ib}}}$, as long as we use the
transformed contact term. Applied to the contact term, the unitaries turn $\gamma _1$ into a linear combination of the other Majorana's in the chain:
\be
\prodd_j e^{iS_j}\gamma_{lead}\gamma _1\prodd_j e^{-iS_j}= \summ_{n~odd}a_{n} \gamma_{lead}\gamma_{n}
\label{lincomb}
\ee
The bipartite geometry of the chain guarantees that only odd
sites can couple to the lead. At the end of the RG
process, the STM lead effectively couples to a Majorana site in many
decoupled Majorana pairs, and the total current is the sum of the
current into each pair (Eq. \ref{LDOS}).  The dominant
contributions, however, arise when the active edge of the chain (at
any stage of the RG) are decimated. E.g., consider the leftmost
Majorana, $\gamma_1$. Its decimation due to a strong bond with
$\gamma_2$, leads to
\be
e^{iS}\gamma_1e^{-iS}=\gamma_1+\frac{t_2}{2t_1}\gamma_3.
\label{edgeex}
\ee 
Subsequent decimations of the new edge $\gamma_3$ couples the tip to
the next odd-numbered Majorana site which is the effective left edge
at that energy scale. Bulk decimations not containing
the edge produce only subdominant correlations contributions in
Eq. (\ref{lincomb}). The tunneling $dI/dV$ will have
strong spikes corresponding to the edge decimations, between which,
bulk decimations yield additional spikes of various
strengths (but always weaker than both edge spikes; see below).

Next, we find the probability distribution for the coupling
coefficients $a_n$, and the energies of the Majorana-pairs they
connect to the edge. These are obtained from the universal coupling
distributions of the effective bond strength at low energies. Define
$\Omega={\rm max}\{t_i\}$ as the maximum coupling in the chain at a
given point in the RG process.  Also, define the logarithmic bond
variables: $\zeta_i=\ln\Omega/t_i$, and logarithmic flow parameter
$\Gamma=\ln \Omega_I/\Omega$, with $\Omega_I$ the largest initial
coupling.  After many
bond decimations, the distribution of bonds settles into the universal
random-singlet fixed point distribution:~\cite{DSF94,DSF95}
\be
P(\zeta)=\frac{1}{\Gamma}e^{-\zeta/\Gamma}.
\ee

We define $u_n(\tau; \Gamma)$ as the probability that when the RG flow
parameter reaches $\Gamma$, the active edge Majorana is coupled with
amplitude $e^{-\tau}$ to the lead, and is the result of  n  previous
decimations involving the edge Majorana, as outlined in
Eq. (\ref{edgeex}).  The evolution equation for $u_n$ is given by
$\frac{du_n}{d\Gamma}=$
\be
-P_0u_n(\tau)+\int d\tau' d\zeta_2 P_0
u_{n-1}(\tau') P(\zeta_2)\delta(\tau-\tau'-\zeta _2),
\label{evolveeq}
\ee
where $P_0=P(0)$, and the probability of a bond to be decimated as
$\Gamma$ changes by $d\Gamma$ is $P_0d\Gamma$. The first term in
Eq.~\ref{evolveeq} marks the decrease in this probability as the
edge is decimated.  The
second is a source term for the $n$'th edge, passing the correlation
information from the edge after $n-1$ decimations. The delta function
expresses the reduction in 
correlation of the new edge compared to its predecessor, by a factor
$e^{-\zeta_2}=t_2/\Omega$ [as in Eq. (\ref{edgeex})].  $\Omega=t_1$ since the
first bond is decimated, and a factor of 2 was neglected since to
logarithmic accuracy we expect $\zeta_2\gg \ln(2)$
(see, e.g.,~\cite{DSF98}). The solution to Eq.~\ref{evolveeq} is 
\beq
u_n = \frac{\Gamma _0}{\Gamma}\frac{1}{n!}\ln ^n \left( \frac{\lambda + 1/\Gamma _0}{\lambda + 1/\Gamma}\right)
\label{un}
\eeq
where $\lambda$ is the Laplace transform variable of $\tau$. 

From $u_n$ it is easy to state the probability density for the $n$'th edge to be decimated at $\Gamma$ and with correlation $e^{-\tau}$ to the edge. It is simply
\be
dJ_n^{(edge)}/d\Gamma = P_0 u_n=\frac{1}{\Gamma}u_n
\label{dJedge}
\ee
As the voltage is scanned from the band-edge towards the band center, the edge decimation will produce $dI/dV$ peaks at values of $V_n$ with a distribution in $\Gamma_n=\ln \Omega_I/V_n$ given by $dJ_n^{(edge)}/d\Gamma$ at $\lambda=0$. Finding the maximum of this distribution yields the most probable voltage of the $n$'th edge decimation:
\beq
V_n^{max}=E_n^{(max)} \sim e^{-e^{n/2}}.
\label{peakloca}
\eeq

\begin{figure}
\includegraphics[width=0.4\textwidth]{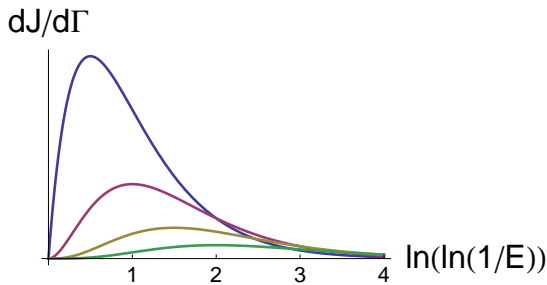}
\caption{\label{locations} Marginal distribution of edge energies for $n$ = 1-4.  The $n$'th peak has a maximum at $\ln (\ln(1/E))$ = n/2.  $dI/dV$ thus has peaks at regular intervals in $\ln(\ln(1/V))$.}
\end{figure}

The statistical properties of the amplitude of the $n$'th $dI/dV$
peak, which is $w_n=t_0^2 e^{-2\tau}$, are also encoded in
$dJ_n^{(edge})/d\Gamma(\tau,\Gamma)$. For instance, the average height
of the $n$'th peak, $\langle e^{-2\tau _n ^{(edge)}}\rangle$, is given
by integrating over $\Gamma$ in Eq. (\ref{dJedge}) to obtain the transform of the marginal distribution for $\tau$, and then setting $\lambda=2$. This yields, roughly, $\langle w_n\rangle\sim \frac{1}{n!}$. This average, however, is dominated by samples with anomalously high correlations. More useful is the typical weight of the $n$'th edge peak, $w_n^{(edge)}=e^{-2\langle \tau \rangle ^{(edge)}_n}$, given that the energy of the $n$'th peak is $V_n^{(max)}$. This yields the behavior:
\beq
w_n^{typial}  \sim e^{-4\Gamma _0 e^{n/2}}
\label{peaksize}
\eeq
It can be shown~\cite{vsforthcoming} that summing up all currents due
to bulk decimations occuring between the $n$'th and $n+1$'th edge
peaks gives $w_n^{(bulk)} \sim w_n ^{(edge)} e^{-3\Gamma _0 e^{n/2}}$,
which allows us to safely neglect them in the single-lead setup.
 
\begin{figure}
\includegraphics[width=0.4\textwidth]{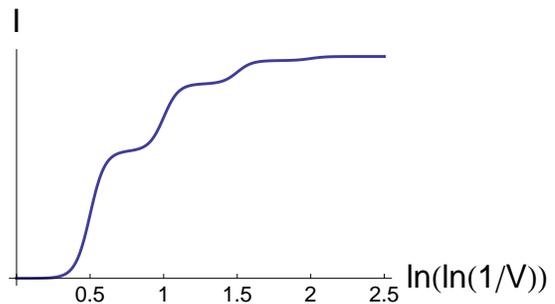}
\caption{\label{smoothIV} Typical IV curve obtained by integrating $dI/dV$ from high voltage down to $V$.  A sharp rise in current occurs at values of $V$ corresponding to typical energies of edge pairs.} 
\end{figure}

Fig.~\ref{smoothIV} shows the predicted current plotted vs.
$\ln(\ln(1/V))$.  Each time the voltage crosses an edge peak from
above, there is a sharp rise in the current as the probe couples to a
decimated Majorana pair.  For higher $n$, i.e., edge decimations at
lower energies, the typical coupling and current jumps decrease sharply.

In this manuscript we calculated the distributions of the STM
tunneling characteristic of an experimentally feasible realization of
an interacting Majorana chain. The observation of the
unusual tunneling STM current we find would be strong evidence
for the critical Majorana chain. The detailed transport information
was obtained using a unitary transformations version of the
Ma-Dasgupta RSRG procedure, which allows going beyond previous studies of
transport in random-singlet type critical points, where only averages
could be obtained~\cite{damlemotrunich}.  This technique could also be
applied to the edge tunneling conductance of the random critical point
of an ordinary fermionic bipartite chain. In future
work we will explore this spectrum in more detail, as well as consider
2-tip transport through a Majorana chain.

The authors thank  L. Fu, C. L. Kane, and L. Molenkamp for useful
conversations and acknowledge support from NSF DMR-0804413 (V.S. and
J.E.M.) and from the Packard Foundation, The Sloan Foundation, the Research
Corporation, and DARPA (G.R).

\bibliography{bigbib}

\begin{thebibliography}{25}
\expandafter\ifx\csname natexlab\endcsname\relax\def\natexlab#1{#1}\fi
\expandafter\ifx\csname bibnamefont\endcsname\relax
  \def\bibnamefont#1{#1}\fi
\expandafter\ifx\csname bibfnamefont\endcsname\relax
  \def\bibfnamefont#1{#1}\fi
\expandafter\ifx\csname citenamefont\endcsname\relax
  \def\citenamefont#1{#1}\fi
\expandafter\ifx\csname url\endcsname\relax
  \def\url#1{\texttt{#1}}\fi
\expandafter\ifx\csname urlprefix\endcsname\relax\def\urlprefix{URL }\fi
\providecommand{\bibinfo}[2]{#2}
\providecommand{\eprint}[2][]{\url{#2}}

\bibitem[{\citenamefont{Fu and Kane}()}]{fukaneprox2d}
\bibinfo{author}{\bibfnamefont{L.}~\bibnamefont{Fu}} \bibnamefont{and}
  \bibinfo{author}{\bibfnamefont{C.~L.} \bibnamefont{Kane}},
  \bibinfo{howpublished}{cond-mat/0804.4469}.

\bibitem[{\citenamefont{Fu and Kane}(2008)}]{fukaneprox3d}
\bibinfo{author}{\bibfnamefont{L.}~\bibnamefont{Fu}} \bibnamefont{and}
  \bibinfo{author}{\bibfnamefont{C.~L.} \bibnamefont{Kane}},
  \bibinfo{journal}{Phys. Rev. Lett.} \textbf{\bibinfo{volume}{100}},
  \bibinfo{pages}{096407} (\bibinfo{year}{2008}).

\bibitem[{\citenamefont{Nilsson et~al.}(2008)\citenamefont{Nilsson, Akhmerov,
  and Beenakker}}]{nilsson}
\bibinfo{author}{\bibfnamefont{J.}~\bibnamefont{Nilsson}},
  \bibinfo{author}{\bibfnamefont{A.~R.} \bibnamefont{Akhmerov}},
  \bibnamefont{and} \bibinfo{author}{\bibfnamefont{C.~W.~J.}
  \bibnamefont{Beenakker}}, \bibinfo{journal}{Phys. Rev. Lett.}
  \textbf{\bibinfo{volume}{101}}, \bibinfo{pages}{120403}
  (\bibinfo{year}{2008}).

\bibitem[{\citenamefont{Kane and Mele}(2005)}]{km2}
\bibinfo{author}{\bibfnamefont{C.~L.} \bibnamefont{Kane}} \bibnamefont{and}
  \bibinfo{author}{\bibfnamefont{E.~J.} \bibnamefont{Mele}},
  \bibinfo{journal}{Phys. Rev. Lett.} \textbf{\bibinfo{volume}{95}},
  \bibinfo{pages}{226801} (\bibinfo{year}{2005}).

\bibitem[{\citenamefont{{Bernevig} et~al.}(2006)\citenamefont{{Bernevig},
  {Hughes}, and {Zhang}}}]{zhangscience1}
\bibinfo{author}{\bibfnamefont{B.~A.} \bibnamefont{{Bernevig}}},
  \bibinfo{author}{\bibfnamefont{T.~L.} \bibnamefont{{Hughes}}},
  \bibnamefont{and} \bibinfo{author}{\bibfnamefont{S.-C.}
  \bibnamefont{{Zhang}}}, \bibinfo{journal}{Science}
  \textbf{\bibinfo{volume}{314}}, \bibinfo{pages}{1757} (\bibinfo{year}{2006}).

\bibitem[{\citenamefont{{Koenig} et~al.}(2007)\citenamefont{{Koenig},
  {Wiedmann}, {Bruene}, {Roth}, {Buhmann}, {Molenkamp}, {Qi}, and
  {Zhang}}}]{molenkampscience}
\bibinfo{author}{\bibfnamefont{M.}~\bibnamefont{{Koenig}}},
  \bibinfo{author}{\bibfnamefont{S.}~\bibnamefont{{Wiedmann}}},
  \bibinfo{author}{\bibfnamefont{C.}~\bibnamefont{{Bruene}}},
  \bibinfo{author}{\bibfnamefont{A.}~\bibnamefont{{Roth}}},
  \bibinfo{author}{\bibfnamefont{H.}~\bibnamefont{{Buhmann}}},
  \bibinfo{author}{\bibfnamefont{L.~W.} \bibnamefont{{Molenkamp}}},
  \bibinfo{author}{\bibfnamefont{X.-L.} \bibnamefont{{Qi}}}, \bibnamefont{and}
  \bibinfo{author}{\bibfnamefont{S.-C.} \bibnamefont{{Zhang}}},
  \bibinfo{journal}{Science} \textbf{\bibinfo{volume}{318}},
  \bibinfo{pages}{766} (\bibinfo{year}{2007}).

\bibitem[{\citenamefont{Moore and Balents}(2007)}]{moore&balents-2006}
\bibinfo{author}{\bibfnamefont{J.~E.} \bibnamefont{Moore}} \bibnamefont{and}
  \bibinfo{author}{\bibfnamefont{L.}~\bibnamefont{Balents}},
  \bibinfo{journal}{Phys. Rev. B} \textbf{\bibinfo{volume}{75}},
  \bibinfo{pages}{121306} (\bibinfo{year}{2007}).

\bibitem[{\citenamefont{Roy}(2009)}]{rroy3D}
\bibinfo{author}{\bibfnamefont{R.}~\bibnamefont{Roy}}, \bibinfo{journal}{Phys.
  Rev. B} \textbf{\bibinfo{volume}{79}}, \bibinfo{pages}{195322}
  (\bibinfo{year}{2009}).

\bibitem[{\citenamefont{Fu et~al.}(2007)\citenamefont{Fu, Kane, and
  Mele}}]{fu&kane&mele-2007}
\bibinfo{author}{\bibfnamefont{L.}~\bibnamefont{Fu}},
  \bibinfo{author}{\bibfnamefont{C.~L.} \bibnamefont{Kane}}, \bibnamefont{and}
  \bibinfo{author}{\bibfnamefont{E.~J.} \bibnamefont{Mele}},
  \bibinfo{journal}{Phys. Rev. Lett.} \textbf{\bibinfo{volume}{98}},
  \bibinfo{pages}{106803} (\bibinfo{year}{2007}).

\bibitem[{\citenamefont{Hsieh et~al.}(2008)\citenamefont{Hsieh, Qian, Wray,
  Xia, Hor, Cava, and Hasan}}]{hsieh}
\bibinfo{author}{\bibfnamefont{D.}~\bibnamefont{Hsieh}},
  \bibinfo{author}{\bibfnamefont{D.}~\bibnamefont{Qian}},
  \bibinfo{author}{\bibfnamefont{L.}~\bibnamefont{Wray}},
  \bibinfo{author}{\bibfnamefont{Y.}~\bibnamefont{Xia}},
  \bibinfo{author}{\bibfnamefont{Y.~S.} \bibnamefont{Hor}},
  \bibinfo{author}{\bibfnamefont{R.~J.} \bibnamefont{Cava}}, \bibnamefont{and}
  \bibinfo{author}{\bibfnamefont{M.~Z.} \bibnamefont{Hasan}},
  \bibinfo{journal}{Nature} \textbf{\bibinfo{volume}{452}},
  \bibinfo{pages}{970} (\bibinfo{year}{2008}).

\bibitem[{\citenamefont{Read and Green}(2000)}]{readgreen}
\bibinfo{author}{\bibfnamefont{N.}~\bibnamefont{Read}} \bibnamefont{and}
  \bibinfo{author}{\bibfnamefont{D.}~\bibnamefont{Green}},
  \bibinfo{journal}{Phys. Rev. B} \textbf{\bibinfo{volume}{61}},
  \bibinfo{pages}{10267} (\bibinfo{year}{2000}).

\bibitem[{\citenamefont{Kitaev}()}]{kitaev2000}
\bibinfo{author}{\bibfnamefont{A.}~\bibnamefont{Kitaev}},
  \bibinfo{howpublished}{cond-mat/0010440}.

\bibitem[{\citenamefont{Bonesteel and Yang}(2007)}]{bonesteelyang}
\bibinfo{author}{\bibfnamefont{N.~E.} \bibnamefont{Bonesteel}}
  \bibnamefont{and} \bibinfo{author}{\bibfnamefont{K.}~\bibnamefont{Yang}},
  \bibinfo{journal}{Phys. Rev. Lett.} \textbf{\bibinfo{volume}{99}},
  \bibinfo{pages}{140405} (\bibinfo{year}{2007}).

\bibitem[{\citenamefont{Bolech and Demler}(2007)}]{bolechdemler}
\bibinfo{author}{\bibfnamefont{C.~J.} \bibnamefont{Bolech}} \bibnamefont{and}
  \bibinfo{author}{\bibfnamefont{E.}~\bibnamefont{Demler}},
  \bibinfo{journal}{Phys. Rev. Lett.} \textbf{\bibinfo{volume}{98}},
  \bibinfo{pages}{237002} (\bibinfo{year}{2007}).

\bibitem[{\citenamefont{Tanaka et~al.}(2009)\citenamefont{Tanaka, Yokoyama, and
  Nagaosa}}]{tanakaprl}
\bibinfo{author}{\bibfnamefont{Y.}~\bibnamefont{Tanaka}},
  \bibinfo{author}{\bibfnamefont{T.}~\bibnamefont{Yokoyama}}, \bibnamefont{and}
  \bibinfo{author}{\bibfnamefont{N.}~\bibnamefont{Nagaosa}},
  \bibinfo{journal}{Phys. Rev. Lett.} \textbf{\bibinfo{volume}{103}},
  \bibinfo{pages}{107002} (\bibinfo{year}{2009}).

\bibitem[{\citenamefont{Asano et~al.}()\citenamefont{Asano, Tanaka, and
  Nagaosa}}]{asanotanaka}
\bibinfo{author}{\bibfnamefont{Y.}~\bibnamefont{Asano}},
  \bibinfo{author}{\bibfnamefont{Y.}~\bibnamefont{Tanaka}}, \bibnamefont{and}
  \bibinfo{author}{\bibfnamefont{N.}~\bibnamefont{Nagaosa}},
  \bibinfo{howpublished}{cond-mat/10044092}.

\bibitem[{\citenamefont{Xu and Fu}()}]{xufu}
\bibinfo{author}{\bibfnamefont{C.}~\bibnamefont{Xu}} \bibnamefont{and}
  \bibinfo{author}{\bibfnamefont{L.}~\bibnamefont{Fu}},
  \bibinfo{howpublished}{cond-mat/0911.1782}.

\bibitem[{\citenamefont{Fisher}(1995)}]{DSF95}
\bibinfo{author}{\bibfnamefont{D.~S.} \bibnamefont{Fisher}},
  \bibinfo{journal}{Phys. Rev. B} \textbf{\bibinfo{volume}{51}},
  \bibinfo{pages}{6411} (\bibinfo{year}{1995}).

\bibitem[{\citenamefont{Refael and Moore}(2004)}]{refaelmoore}
\bibinfo{author}{\bibfnamefont{G.}~\bibnamefont{Refael}} \bibnamefont{and}
  \bibinfo{author}{\bibfnamefont{J.~E.} \bibnamefont{Moore}},
  \bibinfo{journal}{Phys. Rev. Lett.} \textbf{\bibinfo{volume}{93}},
  \bibinfo{eid}{260602} (\bibinfo{year}{2004}).

\bibitem[{\citenamefont{Fisher}(1994)}]{DSF94}
\bibinfo{author}{\bibfnamefont{D.~S.} \bibnamefont{Fisher}},
  \bibinfo{journal}{Phys. Rev. B} \textbf{\bibinfo{volume}{50}},
  \bibinfo{pages}{3799} (\bibinfo{year}{1994}).

\bibitem[{\citenamefont{Hyman et~al.}(1996)\citenamefont{Hyman, Yang, Bhatt,
  and Girvin}}]{GirvinYang}
\bibinfo{author}{\bibfnamefont{R.~A.} \bibnamefont{Hyman}},
  \bibinfo{author}{\bibfnamefont{K.}~\bibnamefont{Yang}},
  \bibinfo{author}{\bibfnamefont{R.~N.} \bibnamefont{Bhatt}}, \bibnamefont{and}
  \bibinfo{author}{\bibfnamefont{S.~M.} \bibnamefont{Girvin}},
  \bibinfo{journal}{Phys. Rev. Lett.} \textbf{\bibinfo{volume}{76}},
  \bibinfo{pages}{839} (\bibinfo{year}{1996}).

\bibitem[{\citenamefont{Refael and Fisher}(2004)}]{Refael2003}
\bibinfo{author}{\bibfnamefont{G.}~\bibnamefont{Refael}} \bibnamefont{and}
  \bibinfo{author}{\bibfnamefont{D.~S.} \bibnamefont{Fisher}},
  \bibinfo{journal}{Phys. Rev. B} \textbf{\bibinfo{volume}{70}}
  (\bibinfo{year}{2004}).

\bibitem[{\citenamefont{Fisher and Young}(1998)}]{DSF98}
\bibinfo{author}{\bibfnamefont{D.~S.} \bibnamefont{Fisher}} \bibnamefont{and}
  \bibinfo{author}{\bibfnamefont{A.~P.} \bibnamefont{Young}},
  \bibinfo{journal}{Phys. Rev. B} \textbf{\bibinfo{volume}{58}},
  \bibinfo{pages}{9131} (\bibinfo{year}{1998}).

\bibitem[{\citenamefont{Shivamoggi et~al.}()\citenamefont{Shivamoggi, Refael,
  and Moore}}]{vsforthcoming}
\bibinfo{author}{\bibfnamefont{V.}~\bibnamefont{Shivamoggi}},
  \bibinfo{author}{\bibfnamefont{G.}~\bibnamefont{Refael}}, \bibnamefont{and}
  \bibinfo{author}{\bibfnamefont{J.~E.} \bibnamefont{Moore}},
  \bibinfo{howpublished}{to be published}.

\bibitem[{\citenamefont{Motrunich et~al.}(2001)\citenamefont{Motrunich, Damle,
  and Huse}}]{damlemotrunich}
\bibinfo{author}{\bibfnamefont{O.}~\bibnamefont{Motrunich}},
  \bibinfo{author}{\bibfnamefont{K.}~\bibnamefont{Damle}}, \bibnamefont{and}
  \bibinfo{author}{\bibfnamefont{D.~A.} \bibnamefont{Huse}},
  \bibinfo{journal}{Phys. Rev. B} \textbf{\bibinfo{volume}{63}},
  \bibinfo{pages}{134424} (\bibinfo{year}{2001}).

\end{thebibliography}
\end{document}